\documentstyle[prb,preprint,aps]{revtex}

\begin{document}
\input{psfig}

\title
{\bf Quantum Computatinal Gates with Radiation Free Couplings}

\author{ I. O. Kulik$^1$, T. Hakio\u{g}lu$^1$ and A. Barone$^2$}
\address{
$^1$Department of Physics, Bilkent University, Ankara 06533, Turkey\\
$^2$Department of Physical Sciences, University of Naples Federico II,
P. Tecchio 80, Naples 80125, Italy\\}

\maketitle
\begin{abstract}
We examine a generic three state mechanism which realizes all
fundamental single and double qubit quantum logic gates operating under
the effect of adiabatically controllable  
static (radiation free) bias couplings between the states. At the instant 
of time 
that the gate operations are defined the third level is unoccupied which, 
in a certain sense derives analogy with the recently suggested dissipation 
free qubit subspaces. The physical implementation of the mechanism is 
tentatively suggested in a form of the Aharonov-Bohm persistent-current 
loop in crossed electric and magnetic fields, with the output of the loop 
read out by a (quantum) Hall effect aided mechanism. \\   
\end{abstract}

{\bf {\em Introduction}}  
Quantum computation is based on the realization of the computational
logic gates by the manipulation of the quantum states via turning on and
off specific interactions controlled by quantum switches.
Recent trends in the experimentation of the small scale quantum logic gates
suggested for implementation in the developing quantum
computers \cite{Nielsen} demands effective gate operations to be performed
by more efficient mechanisms based on dissipation free interactions
between the qubit states that are designed to reduce environment induced
decoherence effects.
There are fundamental differences between the well explored classical and the
much less explored quantum computation and this primarily arises from
the manifestations of the fundamental principles of quantum
mechanics.
From the information theory point of view, quantum computation makes full use
of such classically peculiar concepts as the superposition and entanglement
principles to its maximal advantage to create what is often called the
principle of quantum parallelism via which exponentially unsurpassed
performances are expected in tackling special algorithmic problems such
as the quantum factorization algorithm due to Shor\cite{Shor}, the Grover's
unstructured data search\cite{Grover}, or, more generally, quantum simulation
of the many-body problems (e.g. Ref.\,[4]).
On the physical point of view, making full use of
quantum mechanics in a computational frame requires a full control of the
interacting quantum system with the {\it external} classical system including
the input-output measurement devices as well as not so fully controllable
environmental agents. The delicate susceptibility of the supposedly
isolated evolution of the interacting
quantum system with the interfering environment as well as the systems own
inherent
fluctuations lead to the unwanted decoherence effects. Most crudely speaking,
decoherence can be summarized in the computational terminology as the loss
of computational information stored in the parameters of the quantum state.
In turn, decoherence leaves a small room both spatially and temporally
to manipulate the quantum system for computational purposes  
and the conditions to fight decoherence are very severe.
In this context, inventing new mechanisms
and designing new experimental systems aiming to minimize
all sources of decoherence is a major task of the research efforts in the
physical aspects of this field.

The environmental coupling can be generally considered to be weak
but not weaker than a realistic level in which it becomes
difficult to single out the major sources of decoherence. In an
opposite context, the coupling to the environment has been
suggested to keep the decoherence under control. Lately new
theoretical mechanisms based on multi level quantum systems were
proposed in which the environment strongly couples to the high
levels under certain conditions but not to a subspace in a direct
way. It is suggested that a strong environmental coupling can be
used to cage the quantum system in the Hilbert space into this
subspace free of dissipation and therefore these dissipation free
subspaces can provide a shelter away from decoherence where the
quantum computational bits (qubits) can be manipulated. The main
requirement in these theories is the preexistence of the
dissipation free subspaces\cite{DFS}. More recently Beige et
al.\cite{Beige} have examined this idea theoretically in a
multi-atom three state model with doubly degenerate ground state
comprising the mentioned dissipationless subspace where the third
levels of the atoms strongly coupled to each other and to a single
cavity mode, and Zanardi and Rasetti\cite{Rasetti} discussed
earlier a somewhat similar four-state model with a
decoherence-dissipation free subspaces. When all atoms are in the
ground state and the cavity mode is unoccupied, the dissipative
interaction is effectively switched off. As result, if the cavity
field is in the vacuum state the subspace comprising the doubly
degenerate ground state is dissipation free. More recently there
is also evidence that dissipation free subspaces can be physically
realized and experimentally tested\cite{fortunato}.

In this article we propose a deceptively similar three state idea to
Beige et al. with the fundamental
exception that the mechanism utilizes radiation free static couplings
to perform the quantum logic gates. The operations are defined as
static interactions between
the first two levels (qubit) with the third (auxiliary) level
and, the end of the operation is defined as the instant such that the desired
qugate is obtained in the qubit subspace with no probability of
occupation in the auxiliary state. Moreover in the proposed model the 
otherwise independent concepts of qubit and quantum gate are unified 
within the same quantum unit.  
Therefore in our model the leakage of the wavefunction into the
third level is not avoided, on the contrary, the dynamical occupation of the
third level is an essential part of the qubit gate operations.
The advantages of the proposed model are that the wavefunction never 
leakes out of the Hilbert space spanned by the three states and the quantum
gates are obtained via radiationless mechanisms 
as they involve time-independent non-resonant interactions. To our
knowledge, similar radiationless qugate operations have not been 
discussed yet. The advantage of the radiationless coupling is clearly  
to suppress substantially 
the otherwise environmental dissipative effects in the case when 
resonant coherent light pulse 
(as in the case of ion trap\cite{CZ} and many other mechanisms) 
or magnetic pulse (as in the case of superconducting systems) are used 
to manipulate the quantum states.     

{\bf{\em Persistent current qubit}}.
The realization of such quantum computation schemes can be
attempted with use of macroscopic quantum interference effects in
superconducting systems (the Josephson effect\cite{Barone}), or
the Aharonov-Bohm\cite{Aharonov} persistent-current states in
nonsuperconducting structures of small
size\cite{Kulik},\cite{Buttiker}. The latter structure naturally
realizes the inverted double-degenerate ground state separated
from the higher energy state(s) by a finite gap and therefore
basically not prone to decoherence. Comparatively to this, the
superconducting junctions in the macroscopic quantum
regime\cite{Orlando} may suffer from the decoherence due to
inavoidable admixture of gapless localized excitations near the
barrier area activated at flip transitions between the degenerate
states (this is seen in the broad resonances of the Schrodinger
Cat states observed 
experimentally\cite{Lukens,Nakamura,Silvestrini}).
In our paper we suggest the persistent current loops for the physical 
realization of qubits and qugates. The three-site loop is  
supplemented by a (macroscopic) nondemolitional measuring device 
(the quantum Hall bar in this case), which performs
both tasks in a coherent, decoherence-free fashion by coupling for a
short time the qubit subspace to the third (auxiliary) level.

The three state system in our consideration is defined to be in a
$\Lambda$-shaped configuration (Fig.1) under zero bias potential, i.e. the
ground state is doubly degenerate and there is a third (auxiliary)
state. One possible realization
 is via a three-sectional mesoscopic ring
intersected by tunnelling barriers (or consisting of overlapping metallic
films separated by thin oxide layers) as shown in Fig.2. The isolated
qubit structure can in principle be realized naturally
as a three-island defect in an
insulating crystal, similar to negative-ion triple vacancy (known as $F_3$
-center) in the alkali halide crystal which can be found in standard
textbooks.\cite{Kittel} The gate manipulations can be performed via an   
Aharonov-Bohm flux perpendicular to the ring together with 
a constant electric field within the plane of the ring (Fig.3).
The information to be implemented into the computational basis of the quantum
computer is stored in the form of amplitudes of the persistent-current states
\cite{Kulik}, \cite{Buttiker} of the normal-state Aharonov-Bohm ring (the qubit),
and processed via the radiation free transitions between the states
in an invariant subspace, effected by the static bias potentials on the sites
of the ring (the qugates).

In the absence of the bias potentials, the
dynamics is governed by the pure tunnelling Hamiltonian 
\begin{equation}
H=-\tau\,\sum_{n=1}^{3}\,(a_n^{\dagger}\,a_{n+1}\,e^{i\alpha}+
a^{\dagger}_{n+1}\,a_{n}\,e^{-i\alpha})
\label{hamilt.1}
\end{equation}
where $\tau$ is a real tunnelling amplitude between the islands
and $\alpha$ is a controllable phase. Eq.\,(\ref{hamilt.1}) is
represented in the diagonal basis by the eigenenergies
$\epsilon_{m}=-2\,\tau\,\cos({2\pi \over 3}m+\alpha)$.  The
eigenenergies form the $\Lambda$ configuration at the symmetric
point $\alpha=\pi/3$ with the energies $(-1,2,-1)\,\tau$ for
$m=0,1,2$ respectively. The three sites interact by a bias {\it
potential} loop with the site {\it potential}
$V_n=V_0\,\cos{2\pi\,n/3}$ where $n=0,1,2$ is the site index. It
is clear that for this choice, the potential can be obtained by a
conservative
 field since the total potential around the loop vanishes, i.e.
$V_0+V_1+V_2=0$. The total Hamiltonian is then the sum of
Eq.\,(\ref{hamilt.1}) and the site-potential which is represented in the
diagonal basis of $H$ by the matrix (in units of $\tau$)
\begin{equation}
H_{d}+H_1(V_0)=\pmatrix{-1 &\nu &\nu\cr
          \nu &2 &\nu\cr
          \nu &\nu &-1
          }
\label{totalhamilt.2}
\end{equation}
where $H_d=diag(-1,2,-1)$ is the Hamiltonian (\ref{hamilt.1}) in
the diagonal form, $\nu=V_0/2\tau$ is the dimensionless
interaction parameter. The proposed mechanism is designed to be
radiation free and the quantum gate operations are performed by
adiabatically tuning the static potentials. The system is prepared
in a particular ground state (no occupation of the auxiliary
level) and the time evolution is continued until the instant
$t=t^*$ at which the auxiliary level cycles back to its initial
configuration. The other parameters are adjusted so that the
desired single-qubit gate is realized at the end of the single
cycle of the auxiliary level. The unitary time evolution at
$t=t^{*}$ is then given by
\begin{equation}
e^{it^*(H_{d}+H_1(V_0))}=\pmatrix{A & 0 & B \cr
                                  0 & X & 0 \cr
                                  C & 0 & D}
\label{time.evolu.3}
\end{equation}
where $A,B,C,D$ are complex, $X$ is a pure phase and, other than
the unitarity condition, no other restrictions apply on the matrix
elements. The form of the unitary matrix in
Eq.\,(\ref{time.evolu.3}) leaves the qubit subspace invariant
irrespectively of the value of $X$ as long as the initial
wavefunction is confined to the same subspace. The instantaneous
vanishing of the certain matrix elements in (\ref{time.evolu.3})
is due to the destructive interference in the transition
amplitudes between the auxiliary level and the qubit subspace.
Using the exact expressions describing the absolute level
amplitudes, it can be inferred that the destructive interference
condition at $t=t^*$ can be satisfied if the transition energies
are commensurate. One way to express this condition is
\begin{equation}
E_3-E_1=K\,(E_2-E_3)~,\qquad K=integer~
\label{commensurate.4}
\end{equation}
where $E_i=E_i(V_0)~, ~~i=1,2,3$ ~~are the eigenenergies of
(\ref{totalhamilt.2}) plotted against $V_0$ in Fig.4. In fact,
Eq.\,(\ref{commensurate.4}) is a condition on the static
potential. Solving the eigenvalues of Eq.\,(\ref{totalhamilt.2})
we find that the potential is allowed to take a discrete set of
values determined by
\begin{equation}
V_0^{(K)}=-{2 \over 3K}[K^2+K+1+(K-1)\sqrt{K^2+4K+1}]~.
\label{disc.pot.5}
\end{equation}

{\bf{\em Qugate operations}}.
We now demonstrate that different values of the integer $K$ performs
different qubit gates. In particular,
among the fundamental single qubit gates the bit flip and the Hadamard-like
gates can be realized by the time evolution of the Hamiltonian alone in
Eq.\,(\ref{totalhamilt.2}) at certain instants and at specifically tuned
values of $V_0$. Among the elementary
qubit gates, the phase gate requires a control on the relative phase between
the degenerate states. In order to induce a relative phase, the otherwise
degenerate states in the qubit subspace
are made nondegenerate by a shift in their eigenlevels by
turning on a degeneracy breaking interaction. This is a relatively well known
method in the case of Aharonov-Bohm rings, for instance, by slightly shifting
the dc-flux away from the value where doubly degenerate configuration is
defined. The net effect of this shift in the adiabatic limit is represented
by a diagonal, degeneracy breaking effective term in the total Hamiltonian
\begin{equation}
H_{2}={\rm diag}(\Delta \epsilon_1,\Delta \epsilon_2, \Delta \epsilon_3)~.
\label{shift.flux.5}
\end{equation}
The diagonal form of Eq.\,(\ref{shift.flux.5}) implies that the
phase shift can be obtained independently from the other gates
since, due to the diagonal form, the time evolution is manifestly
adiabatic. One other advantage in this diagonal form is that the
transformation leaves the qubit subspace invariant and thus it can
be conveniently used for phase correction. We demonstrate below
the realization of the different single qubit operations by a mere
change of the integer $K$ and letting the system time evolve. The
populations of the three eigenstates of the Hamiltonian in
Eq.\,(\ref{hamilt.1})  are plotted in Fig.5 as functions of time
and for $K=1$. The first observation is that the maximal
occupation of the auxiliary state is $20 \%$ of the total unit
probability. At periodic time intervals, of which period $t_1$ is
indicated on the horizontal axis by an arrow, the population in
the auxiliary state vanishes and the wavefunction instantaneously
collapses onto the qubit-subspace non-demolitionally. Hence, at $t=t_1$
the degenerate levels exchange their population. The bit flip
should introduce no relative phase between the qubit states, thus,
one needs to know not the probabilities but the amplitudes. These
can be directly obtained from the unitary time evolution at $K=1$
(which corresponds in Eq.(\ref{totalhamilt.2}) to $V_0^{(1)}=-2$).
Evolving the Hamiltonian in (\ref{totalhamilt.2}) at this configuration
for $t_1=\pi/\sqrt{6}$ (in units of $\hbar/\tau$),
$t_1=\pi/\sqrt{6}$ (in units of $\hbar/\tau$) as
\begin{equation}
\exp\{-it_1 \,\pmatrix{-1 & -1 & -1 \cr
                       -1 &  2 & -1 \cr
                       -1 & -1 & -1 \cr}\}=
                        \pmatrix{0 & 0 & -1 \cr
                                  0 & 1 & 0 \cr
                                 -1 & 0 & 0\cr}
\label{bit.flip.7}
\end{equation}
and ignoring the overall phase, we obtain a bit-flip in the qubit
subspace. The second gate manifested by the commensuration
condition is the Hadamard gate which is obtained at $K=3$. In
Fig.6 the occupation of the states are plotted as functions of
time. The period $t_3$ at which the instantaneous collapse to the
qubit subspace with symmetric occupations occurs is indicated by
an arrow. The unitary matrix that performs this operation is
\begin{equation}
\exp\{-it_3\pmatrix{-1 & V_0^{(3)}/2 & V_0^{(3)}/2 \cr
                       V_0^{(3)}/2 & 2 & V_0^{(3)}/2 \cr
                       V_0^{(3)}/2 & V_0^{(3)}/2 & -1\cr}\} =
{e^{i\alpha} \over \sqrt{2}}\pmatrix{1 & 0 & -i \cr
                     0 & \sqrt{2}e^{i\beta} & 0 \cr
                                      -i & 0 & 1\cr}
\label{hadamard.8}
\end{equation}
where $t_3=\pi/2\,[E_2(V_0^{(3)})-E_3(V_0^{(3)})]=0.7043492$
(in units of $\hbar/\tau$) and
$V_0^{(3)}=-2(13+2\sqrt{22})/9$. The $\alpha$ is an overall phase which is
ignored, and, $\beta$ is the phase of the auxiliary level which is irrelevant
for the qubit subspace. The gate in (\ref{hadamard.8}), after
correcting the phase by a relative phase shift becomes a Hadamard
gate in the qubit subspace.
The phase is corrected by a phase gate which is obtained by
turning off $V_0$ and shifting the levels by $H_2$.
The shifted flux removes
the degeneracy with a net effect implied by $H_2$ and, under the time
evolution, phases between the states are induced without changing the
populations. The time dependence of the
transformation induced by the phase gate is
\begin{equation}
G(\phi)=e^{-i\,t\,(H_{d}+H_2)}=\pmatrix{e^{i\phi_1(t)} & 0 & 0\cr
                                       0 & e^{i\phi_2(t)} & 0\cr
                                       0 & 0 & e^{i\phi_3(t)}\cr}
\label{phase.9}
\end{equation}
The relative phase $(\phi_1-\phi_3)/2=\phi(t)$ applies to the
qubit subspace and the phase induced on the auxiliary level can be
totally ignored. The relative phase correction needed in
Eq.\,(\ref{hadamard.8}) can be achieved by sandwiching it between
the two phase gates $G(\phi=-\pi/4)$. By this demonstration it is
also clear how to perform a phase flip which can be obtained by
producing $\phi=\pi/2$.

The realization of the controlled operations with double qubits is
an essential requirement of any mechanism of quantum computation.
It is possible to obtain a CNOT gate in the quantum system we
propose. Both three level systems are initially prepared to be in
their qubit subspaces and they are connected by a quantum
nondemolitional measurement device which reads the 
first qubit and depending on its state, it induces a static
potential $V_0^{(1)}$ in the second qubit to perform the bit flip.
The experimental scheme is shown in Fig.7 which employs two
mesoscopic rings, a Hall bar in the full quantum regime and
superconducting loop. The flux in the qubit No.1 (which includes
the externally applied flux and the flux created by a persistent
current) is extracted from the former by a $-\Phi_0/2$
compensating coil, and further supplied to the Hall bar with a
(large) current passing through it. The Hall voltage generated in
the bar is designed so that either $V_0^{(1)}$ or zero voltage is
produced corresponding to the fixed value of the current flowing
in one or the other direction. The Hall bar is connected to the
$V$ electrodes of qubit No.2. If the voltage is $V_0^{(1)}$, the
bit flip of the second qubit is realized after time $t_1$  or if
the voltage is zero no change is made. The procedure may in
principle be executed in a totally reversible way if the Hall bar
operates in the manifestly quantum
regime.\\
In more detail, assume that the current $J_Q$
generated by the qubit-1 loop
is $J_Q=J_0$ when the qubit is in the clockwise direction
(the $\vert 0\rangle$ state), and value of the current $J_Q=-J_0$ when the
 direction is counterclockwise (the $\vert 1\rangle$ state).
Shifting the current received from the first qubit by an
additional $J_0$ provided by the "minus" flux ring in Fig.7 the
final current becomes binary (i.e. $0$ or $1$) in units of $J_0$.
The binary current is fed to the Hall bar which is designed to
produce a voltage $V_Q=k*(J_Q+J_0)$. Applying this one to the
qubit No.2 will produce a voltage $2kJ_0=V_0^{(1)}=-2$ (in units
of $T$) at a proper choice of the instrumental parameter $k$
(i.e., the magnitude and the sign of the transport current in the
bar $J_0$). The Hall voltage is then $V_0^{(1)}$ when qubit is
$\vert 0\rangle$, and zero when it is in the state $\vert
1\rangle$.

{\bf{\em Physical implementation}}.
Among other crucial points in the computation, are a reproducible
initialization, storing the information until the final readout,
the decoherence effects,
and an accurate readout which we examine respectively. For the
initialization,  the magnetic flux is shifted adiabatically from
half flux quantum and the system is allowed to relax to the
nondegenerate lowest energy state $\vert 0\rangle$ by spontaneous
emission. We either leave the state there or, by applying a
Hadamard gate, a Schr\"{o}dinger Cat state is obtained which is
conventionally the  initial state in some quantum computing
algorithms, in particular Shor's algorithm for factorizing large
integers\cite{Shor}.

Regarding the decoherence, the gate transformation can be the main
source for loss of quantum information since the qubit idling, due
to a never-decaying property of a persistent current\cite{Kulik},
does not introduce losses and decoherence whereas the gate
manipulation, nevertheless being effected with a time independent
Hamiltonian (\ref{totalhamilt.2}), is a time dependent process
which, when allowed for coupling to the radiative environment,
is a source of decoherence. 
Since our working medium is an electronic field, the main
source of dissipative energy loss is the electric dipole radiation with 
the average intensity (the energy loss per unit time) $w =
2<\vert{\bf\ddot{d}}\vert^2>/3c^3 \sim e^2a^2\omega^4/c^3$ where $a$ is
typical size of the persistent-current loop and $\omega$ the
characteristic frequency of switching of the order of the hopping
amplitude $\tau/\hbar$. Hence, the qugate quality factor
$Q\sim\omega T$ where $T$ is the time of decoherence, turns out to
be of order
\begin{equation}
Q \sim (\frac{e^2/a}{\tau})^2\frac{1}{\alpha^3}
\label{decoherence}
\end{equation}
where $\alpha$ is a fine structure constant $e^2/\hbar c$.
Therefore, for a properly choosen hopping amplitude such that
$\tau \leq e^2/a$, the qubit can support about $10^6$ operations
within the computational cycle.

For parallel readout in a large scale computation some of the
produced data needs to be read in parallel and simultaneously in a
number of qubits. This implies that the information in some qubits
must be stored until all necessary operations are performed and
during this time the qubit subspace should be free of dissipation.
The coherence can be maintained by adopting the $H_1=0, H_2=0$
case as the idling configuration in the doubly degenerate
eigenbasis of $H_0$. The degenerate configuration helps the states
to maintain relative phase coherence. 

In conclusion, we suggested a radiation free mechanism 
whose physical Hamiltonian allows for
coupled qubit/qugate storage and processing of information in a reversible,
scalable way with reduced decoherence effects. The system implementation
remains to be a future task which may become less demanding due to high degree
of flexibility in setup organization regarding in particular the use of
multi-loop qubits and quantum mesoscopic effects other than the
original Aharonov-Bohm one, including the Berry phase and spin-orbit interaction
induced persistent currents.
\newpage

\section*{Figure Captions}

Fig.1: $\Lambda$-shaped level configuration of the persistent-current
normal-state Aharonov-Bohm qubit. 1,2,3 are the eigenstates of Hamiltonian
(1) at $\alpha=\pi/3$ where 1 and 3 are the computational basis (qubit)
registers
$\vert 0\rangle$, ~$\vert 1\rangle$, whereas 2 is the qugate (control)
register $\vert c\rangle$. Arrows show transitions between the degenerate
states effected by the potential biases applied to the ring sites.\\

Fig.2: (a)A sketch of magnetically focused lines of magnetic field
(arrows) of the superconducting fluxon with flux $\Phi_1=hc/2e$ making one
half of the normal-metal flux quantum, $\Phi_0=hc/e$, and effecting the
ring ${\bf R}$ with three normal islands into a $\Lambda$-shaped configuration.
The fluxon $\Phi_1$ is trapped in the opening of superconducting foil
(${\bf S}$) and further compressed by a ferromagnetic crystal  to
fit into the interior of the ring.~(b)Schematic of the multi-ring qubit
arrangement with the sites (circles) on the surface of cylindrical wall
${\bf R}$ surrounding ferromagnetic cylinder (${\bf F}$) which focuses lines
of magnetic field in a cylindrical tube inside superconductor (${\bf S}$).\\

Fig.3: A sketch of the electric field (shown by arrows) applied to the ring through
the potential electrodes (${\bf V}$) in direction perpendicular to the direction
of magnetic field. ${\bf C}$ is a coupling loop providing the connection
to the nondemolition-measuring setup of the persistent current.\\

Fig.4: Eigenenergies of the ring biased with a potential $V_0$. Eigenstates
1 and 3 are degenerate at $V_0=0$ where they form the qubit states. The
commensurate situation, marked by arrows, appears at $K=1$ and at $K=3$ where
it allows for the temporal (virtual) transition to a higher level 2 and back
thus effecting the bit-flip transition (at $K=1$) and the Hadamard-like
gate (at $K=3$).\\

Fig.5: Bit evolution at $K=1$. At point indicated by an arrow ($t=t_1$),
the population of control register ($\vert c\rangle$) vanishes whereas the
populations of the computational registers of qubit ($\vert 0\rangle$) and
($\vert 1\rangle$), interchange.\\

Fig.6: Bit evolution at $K=3$. At point indicated by an arrow ($t=t_3$),
the population of control register ($\vert c\rangle$) vanishes whereas the
computational basis of the qubit, originally in a state ($\vert 1\rangle$,
equally populates to states $\vert 0\rangle$ and $\vert 1\rangle$.\\

Fig.7: A sketch of the ${\bf CNOT}$ quantum gate. The ring No.1
which is pierced by a positive-$\Phi_0/2$ flux (marked ``+''), after
the subtraction of the negative-$\Phi_0/2$ flux (``-''), couples via
the loop ${\bf C}$ to a quantum Hall bar ({\bf H}) through which a
fixed current ({\bf J}) is fed. The bar generates a Hall voltage
output effecting, through the potential electrodes
${\bf V}$, the flip transition in the ring No.2.\\

\newpage

\begin{figure}
\hspace{18mm} \psfig{figure=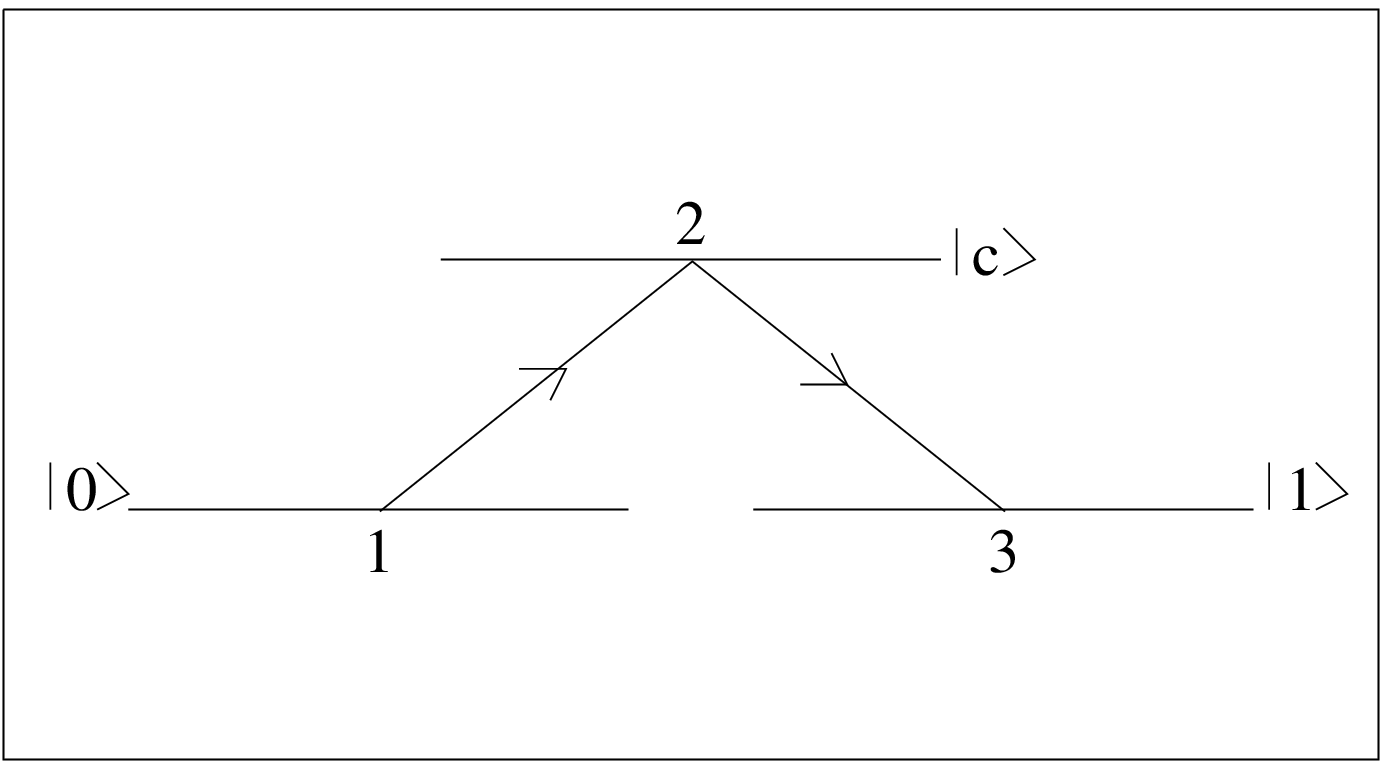,width=9cm,height=5cm}
\end{figure}
\vspace{5mm}
\begin{center}
Fig.1
\end{center}
\newpage

\begin{figure}
\hspace{15mm} \psfig{figure=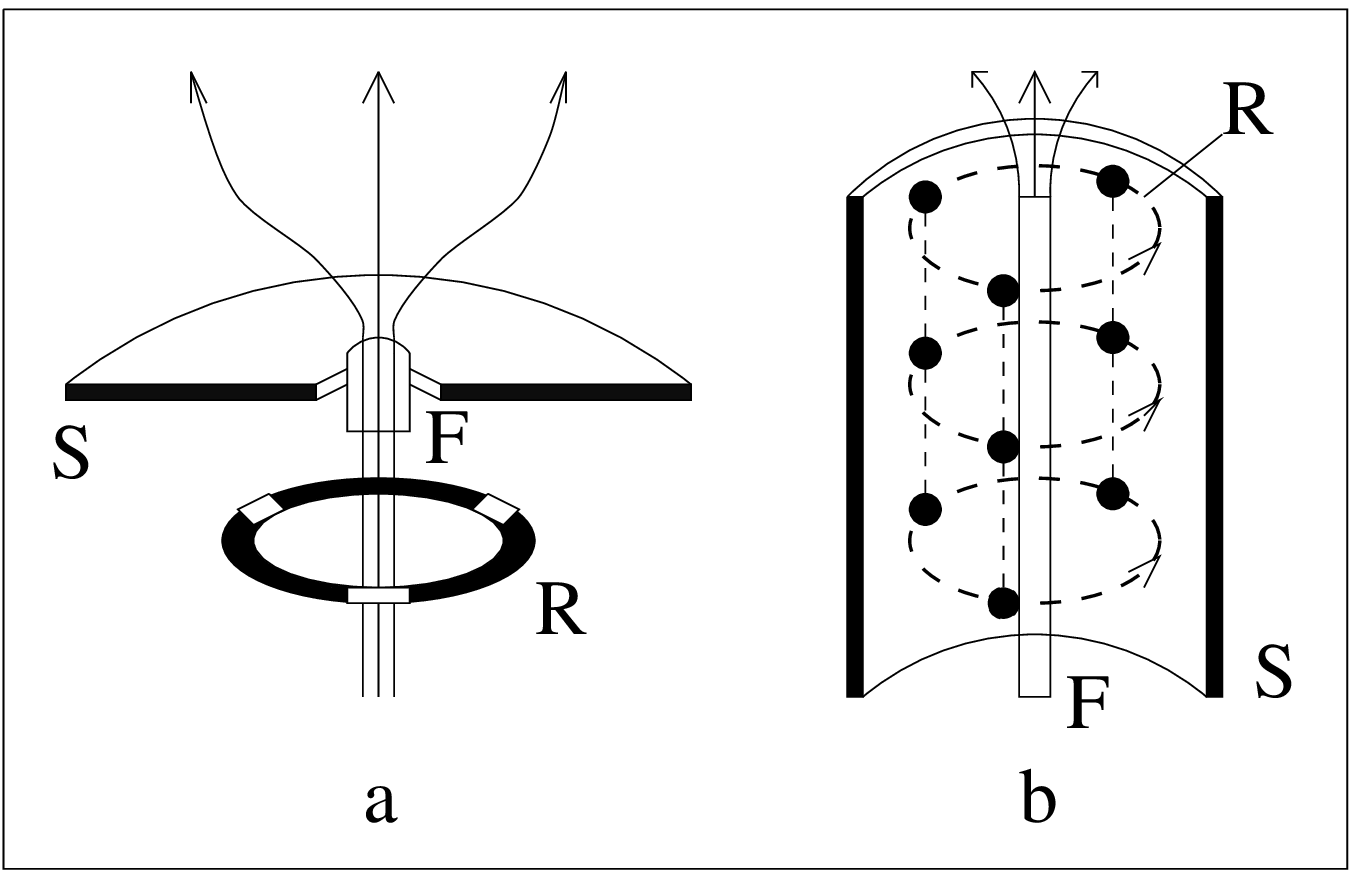,width=12cm,height=8cm}
\end{figure}
\vspace{5mm}
\begin{center}
Fig.2
\end{center}
\newpage

\begin{figure}
\hspace{15mm} \psfig{figure=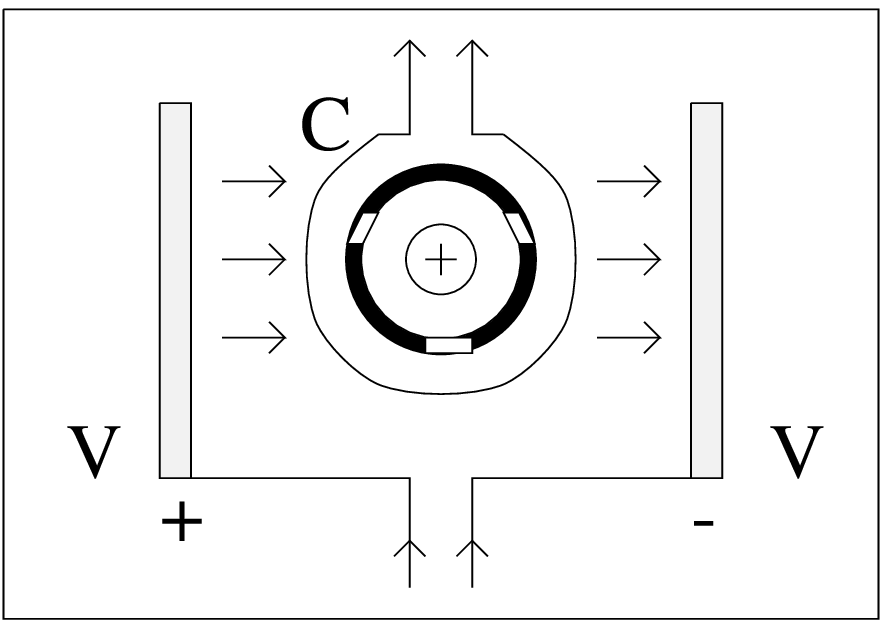,width=10cm,height=7cm}
\end{figure}
\vspace{5mm}
\begin{center}
Fig.3
\end{center}
\newpage

\begin{figure}
\hspace{15mm} \psfig{figure=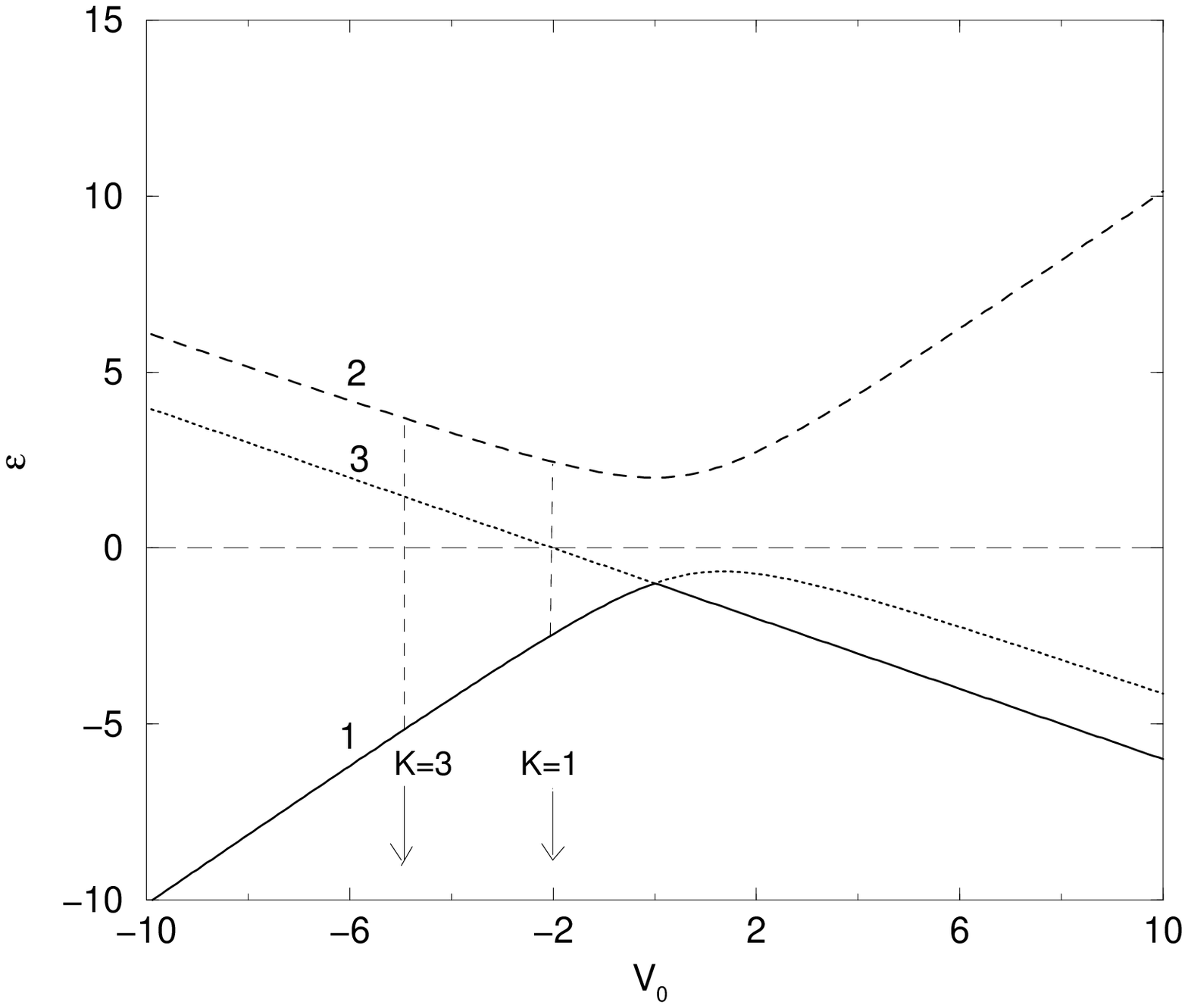,width=12cm,height=8cm}
\end{figure}
\vspace{5mm}
\begin{center}
Fig.4
\end{center}
\newpage

\begin{figure}
\hspace{15mm} \psfig{figure=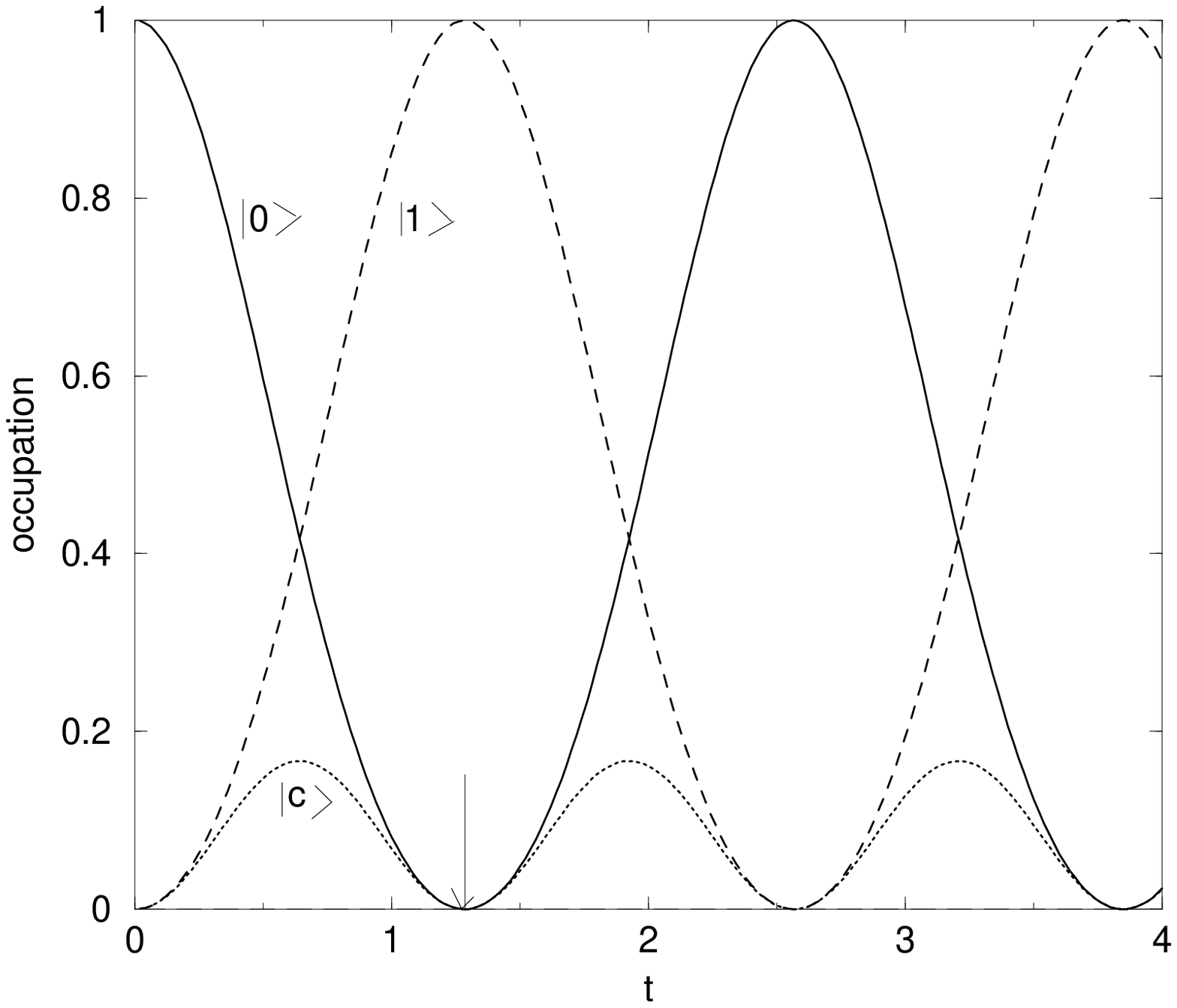,width=12cm,height=8cm}
\end{figure}
\vspace{5mm}
\begin{center}
Fig.5
\end{center}
\newpage

\begin{figure}
\hspace{15mm} \psfig{figure=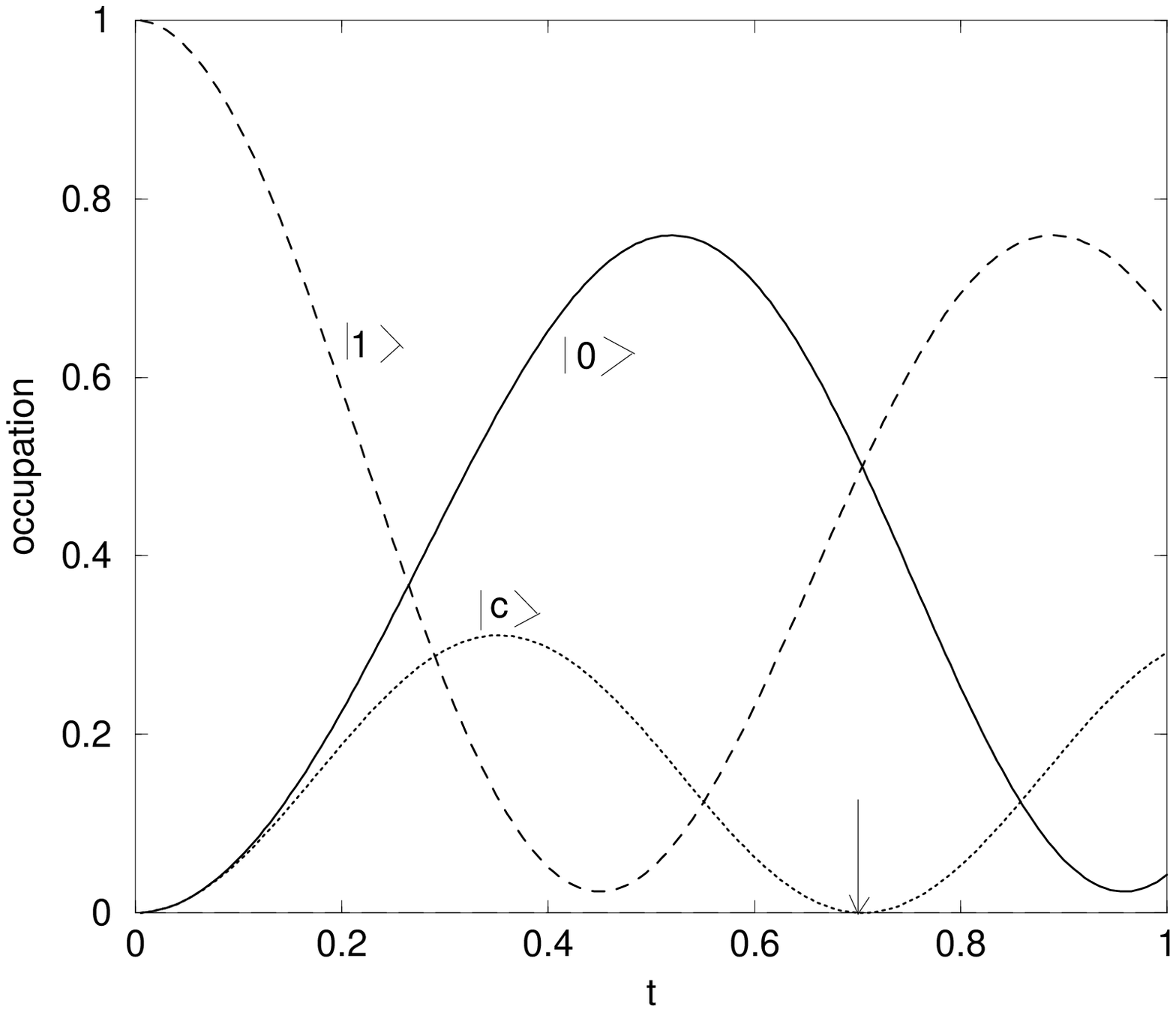,width=12cm,height=8cm}
\end{figure}
\vspace{5mm}
\begin{center}
Fig.6
\end{center}
\newpage

\begin{figure}
\hspace{15mm} \psfig{figure=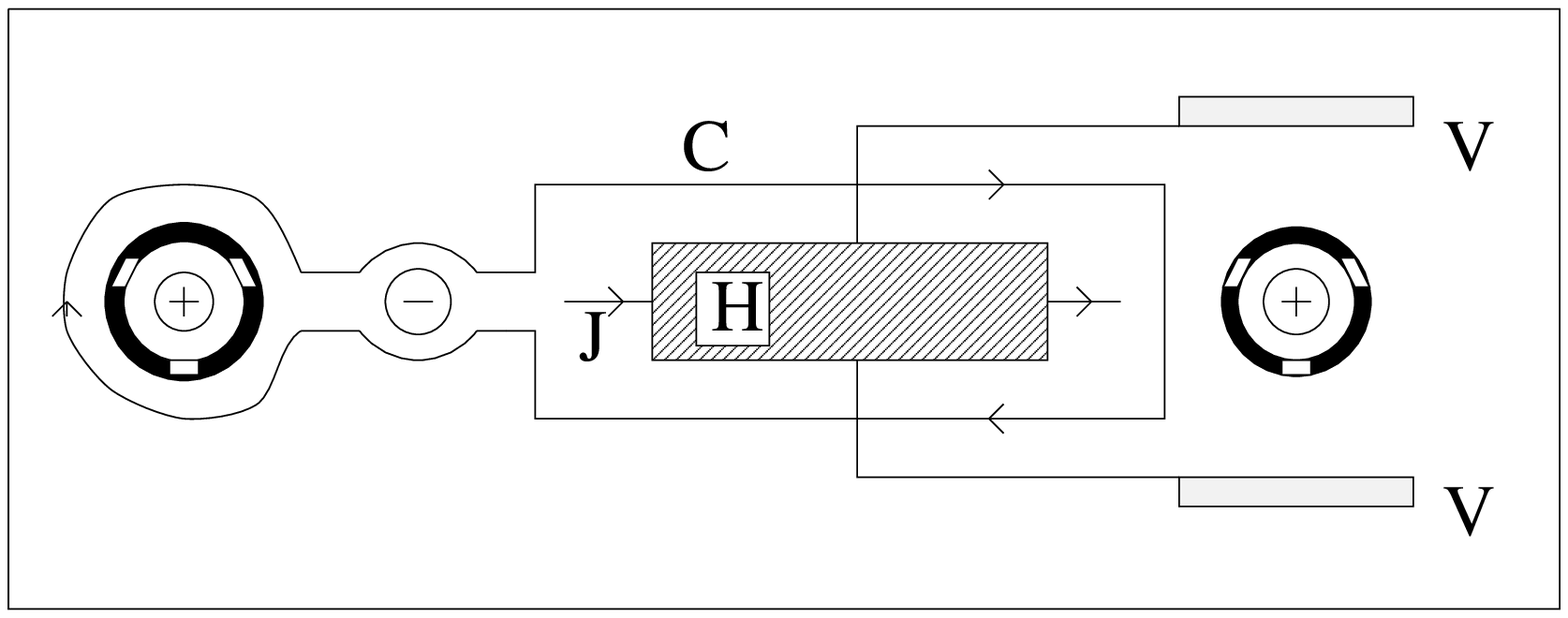,width=13cm,height=5cm}
\end{figure}
\vspace{5mm}
\begin{center}
Fig.7
\end{center}

\end{document}